\documentclass[psfig,preprint,showpacs,apsrev4-1]{revtex4-1}
\usepackage{amsfonts}
\usepackage{amsmath}
\usepackage{graphicx}
\usepackage{bm}
\usepackage{amssymb}
\usepackage{times}
\usepackage{dcolumn}

\begin{document}

\title{Octagraphene as a Versatile Carbon Atomic Sheet for Novel
 Nanotubes, Unconventional Fullerenes and Hydrogen Storage}

\author{Xian-Lei Sheng,$^1$ Hui-Juan Cui,$^1$ Fei Ye,$^2$ Qing-Bo Yan,$^2$ Qing-Rong Zheng $^1$
and Gang Su$^1$\footnote{Corresponding author. E-mail:
gsu@gucas.ac.cn}}

\affiliation{$^1$Theoretical Condensed Matter Physics and
Computational Materials Physics Laboratory, College of Physical
Sciences, Graduate University of Chinese Academy of Sciences, Beijing 100049, China \\
 $^2$College of Materials Science and Opto-Electronic
Technology, Graduate University of Chinese Academy of Sciences,
Beijing 100049, China}

\begin{abstract}

  We study a versatile structurally favorable periodic $sp^2$-bonded carbon
  atomic planar sheet with $C_{4v}$ symmetry by means of the
  first-principles calculations. This carbon allotrope
  is composed of carbon octagons and squares with two bond lengths and
  is thus dubbed as octagraphene. It is a semimetal with the Fermi
  surface consisting of one hole and one electron pocket, whose
  low-energy physics can be well described by a tight-binding model of
  $\pi$-electrons. Its Young's modulus, breaking strength and Poisson's
  ratio are obtained to be 306 $N/m$, 34.4 $N/m$ and 0.13, respectively,
  which are close to those of graphene. The novel sawtooth and armchair
  carbon nanotubes as well as unconventional fullerenes can also be
  constructed from octagraphene. It is found that the Ti-absorbed
  octagraphene can be allowed for hydrogen storage with capacity
  around 7.76 wt\%.

\end{abstract}


\maketitle

\section{Introduction}

The discovery of graphene \cite{geim2004} opens a new era of
fundamental research and technological applications of carbon-based
materials. Graphene is an elegant two-dimensional (2D) material
exhibiting exceptional electrical, thermal, mechanical and optical
properties, and its low-energy behavior of electrons is described by
a (2+1)-dimensional relativistic quantum theory \cite{geim_review,
neto_review,peres_review}. Stimulated by the tremendous interest of
graphene, a great number of novel carbon structures were proposed
and investigated in the past years (e.g. Refs.
[\onlinecite{crespi,chemsocrev,PRB_defect,enyashin,Tcarbon,yakobson}]).
Among others, two intriguing $sp$-$sp^2$-bonded monolayers of
elemental carbon, graphyne \cite{graphyne} and graphdiyne
\cite{graphdiyne,malko}, and an insulating 2D hydrocarbon layer,
graphane \cite{graphanePRB,graphaneEXP}, are typical examples, all
of which were experimentally obtained. It appears that there is a
growingly extensive interest to seek for more 2D allotropes of
carbon with amazing physical and chemical properties as well as
possibly diverse applications.

Graphene has a perfect honeycomb lattice with $C_{6v}$ symmetry.
Recent studies \cite{PRB_defect,line_defect,Cmembrane} show that the
line defects with octagons and pentagons could be formed
self-assembly in graphene. The formation of small carbon molecules
with various polygon rings has also been studied from experimental
aspects \cite{chemrev}. Inspired by these observations, we propose a
stable 2D periodic atomic sheet consisting of carbon octagons,
coined as {\it octagraphene}, which possesses intriguing properties
and might be synthesized experimentally by line defects or acetylene
scaffolding ways. As shown in Fig. 1(a), octagraphene comprises of
carbon octagons and squares with two bond lengths that forms a
square lattice with $C_{4v}$ symmetry. By means of the
first-principles calculations, it is shown that octagraphene is
energetically and kinetically stable, and is more favorable in
energy than graphyne and graphdiyne, although it is metastable
against graphene. By rolling octagraphene along specific directions,
we can get novel sawtooth and armchair carbon nanotubes [Fig. 1(b)]
that are metallic, which can escape from the difficult separation of
usual metallic and semiconducting carbon nanotubes \cite{sorting};
by wrapping octagraphene and making use of pentagons, hexagons or
heptagons as caps, many stable unconventional fullerenes such as
C$_{36}$, C$_{48}$, C$_{80}$, C$_{96}$ and so forth [Fig. 1(c)] can
be constructed. The electronic properties of octagraphene manifest
that it is a semimetal with small density of states (DOS) around the
Fermi level, whose Fermi surface  consists of one hole and one
electron pockets centered at $\Gamma$ and  $M$ points, respectively.
The low-energy physics of octagraphene can  be well described by a
tight-binding model of $\pi$  electrons. Octagraphene has the
density of 0.68 $mg/m^{2}$, Young's  modulus of 306 $N/m$, breaking
strength of 34.4 $N/m$ and Poisson's ratio  of 0.13. It could be the
strongest periodic carbon sheet after graphene till now. An energy
gap can be opened by doping boron nitrogen pairs. The possible
routes for obtaining octagraphene are suggested. The Ti-absorbed
octagraphene can also be used for hydrogen storage with capacity of
7.76 wt\%. With these amazing characters, we anticipate that
octagraphene, similar to its cousin---graphene, once being obtained,
will likewise have a crucial impact in physics, chemistry, materials
and information sciences.

The rest of this article is organized as follows. In Sec. II, the
calculational methods are described. In Sec. III, the geometrical
structure,  mechanical properties, electronic structures and
low-energy physics of octagraphene  are presented. In Sec. IV, new
carbon nanotubes and unconventional  fullerenes from octagraphene
are given. The band engineering and hydrogen storage of octagraphene
are discussed in Sec. V. Finally, a conclusion is briefly presented.

\section{Calculational Methods}

Both the first-principles calculations within the framework of the
density-functional theory (DFT) \cite{dft1,dft2} and the
tight-binding approximation (TBA) \cite{cntbook} are invoked to
study this 2D system. For the first-principles method, primary
calculations were performed within the Vienna \textit{ab initio}
simulation package (VASP) \cite{vasp1,vasp2} with the projector
augmented wave (PAW) method \cite{paw}. Both the local density
approximation (LDA) in the form of Perdew-Zunger \cite{ldapz} and
generalized gradient approximation (GGA) developed by Perdew and
Wang \cite{ggapw} were adopted for the exchange correlation
potentials. The plane-wave cutoff energy is taken as 400 eV. The
Monkhorst-Pack scheme for k-point samplings with
32$\times$32$\times$1 mesh for planar sheets and
1$\times$1$\times$100 mesh for nanotubes were used to sample the
Brillouin zone \cite{MPscheme}. The supercells are used for
calculations of isolated sheet structures, and the distance between
two layers is about 10 {\AA} to avoid interactions. The geometries
were optimized when the remanent Hellmann-Feynman forces on the ions
are less than 0.01 eV/{\AA}. Phonon calculations with a k-mesh of
8$\times$8$\times$1 are performed using the density functional
perturbation theory \cite{espresso}.

\section{Octagraphene}

\begin{figure}[tbp]
\includegraphics[width=10.5cm]{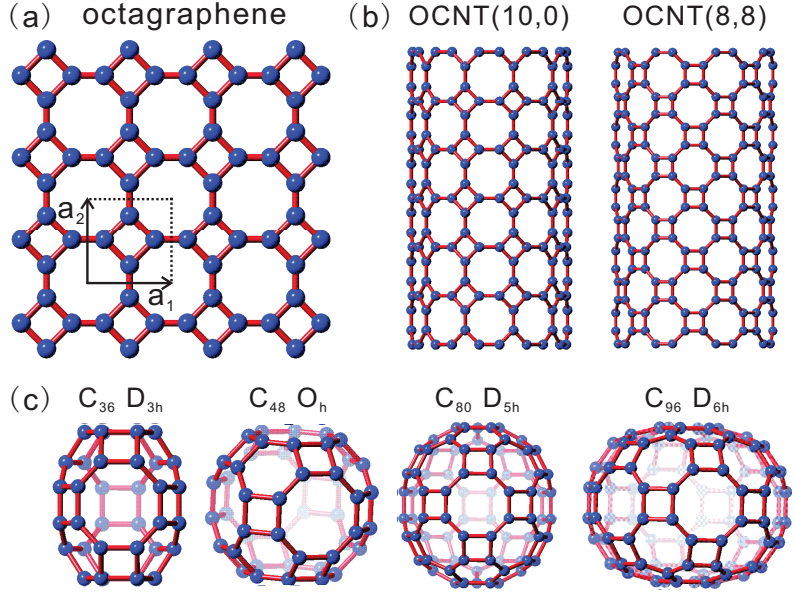}
\caption{(Color online) Schematic depiction of the structures of (a)
octagraphene, where a unit cell is indicated with the unit vectors
$\textbf{a}_{1}$ and $\textbf{a}_{2}$; (b) single-wall sawtooth
(10,0) and armchair (8,8) carbon nanotubes rolled from octagraphene;
and (c) unconventional fullerenes of C$_{36}$, C$_{48}$, C$_{80}$
and C$_{96}$ obtained from octagraphene.}
\end{figure}

\subsection{Geometrical structure and mechanical properties}

The optimized geometric structure of octagraphene is presented in
Fig. 1(a). Carbon atoms in octagraphene form a square lattice with
the lattice constant $a_{0}=3.45$ {\AA}, where a unit cell contains
four carbon atoms. Similar to graphene, every carbon atom in
octagraphene has three $sp^{2}$-bonded nearest neighbors, forming
three $\sigma$ bonds. In addition, octagraphene has two types of
bond lengths, $1.48$ {\AA} of intra-squares and $1.35$ {\AA} of
inter-squares, and two bond angles $90^{\circ}$ and $135^{\circ}$.

\begin{figure}[tbp]
\includegraphics[width=1.0\linewidth,clip]{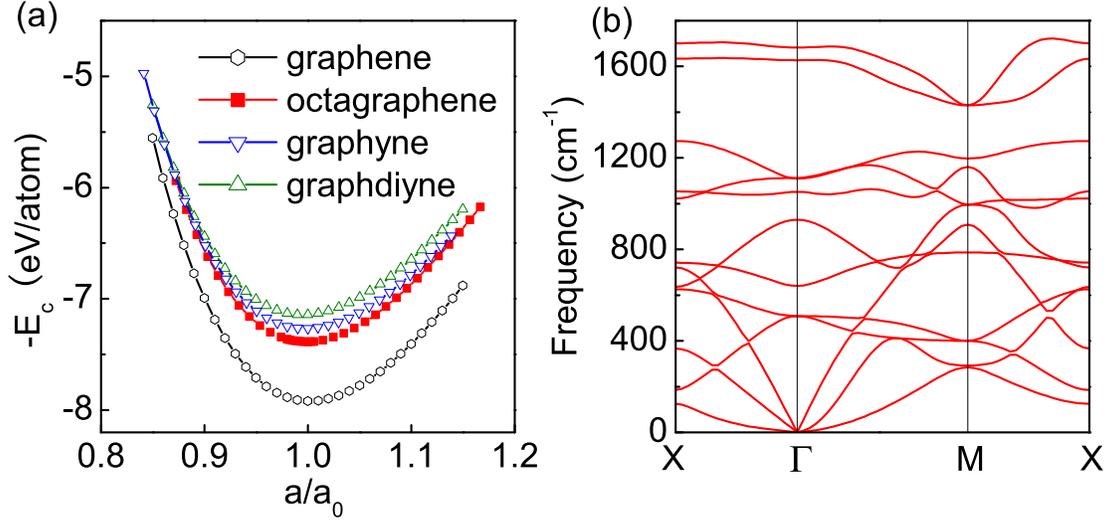}
\caption{(Color online) (a) Cohesive energy per atom as a function
of lattice constant $a$ for graphene, octagraphene, graphyne and
graphdiyne, respectively, where $a_0$ is the optimized lattice
constant. (b) Phonon spectra of octagraphene.}
\end{figure}

The cohesive energy $E_{c}$ per atom as a function of lattice
constant ($a/a_0$) is given in Fig. 2(a) for graphene, octagraphene,
graphyne and graphdiyne, respectively. It is clear that a single
minimum of $-E_{c}$ appears at $a/a_{0}=1$ for these 2D structures,
suggesting that they are energetically stable. Note that the minimum
of $-E_c$ for octagraphene is smaller than those of graphyne and
graphdiyne but larger than that of graphene, suggesting octagraphene
is more energetically stable than graphyne and graphdiyne but
metastable against graphene. Fig. 2(b) gives the phonon spectra of
octagraphene. No imaginary phonon modes are found, implying that
octagraphene is also kinetically stable. We performed quantum
molecular dynamics simulations and found that at temperature 500 K
the planar structure of octagraphene still maintains. Considering
that graphyne and graphdiyne were already obtained experimentally,
it is reasonable to believe that octagraphene can be synthesized in
lab or likely already exists in Nature. In fact, carbon rings with
four or eight sides have already been observed in some experiments
\cite{line_defect,Cmembrane}.

For a comparison, the cohesive energy $E_c$, density $\rho$, bond
length $l_{CC}$, energy gap $E_{g}$, Young's modulus $E$, breaking
strength $\sigma$ and Poisson's ratio $\nu$ for graphene,
octagraphene, graphyne and graphdiyne, respectively, are collected
in Table I. Notice that apart from the experimental data of graphene
taken from literature, all data presented here were calculated by
ourselves using the same method. As octagons in octagraphene are
larger in size than hexagons in graphene, octagraphene has the
density (0.68 $mg/m^2$) slightly smaller than that of graphene (0.77
$mg/m^2$).

\begin{figure}[tbp]
\includegraphics[width=1.0\linewidth,clip]{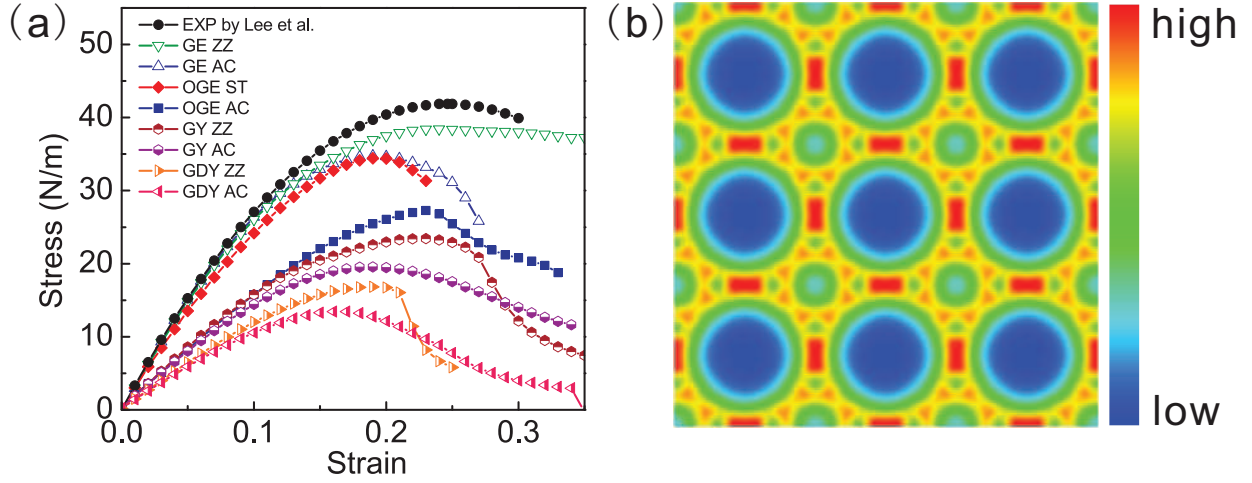}
\caption{(Color online) (a) Stress-strain curves of graphene (GE),
octagraphene (OGE), graphyne (GY) and graphdiyne (GDY) along
armchair (AC) and zigzag (ZZ) or sawtooth (ST) directions. The
experimental data of graphene are taken from \cite{science_tension}.
(b) Valence electron density of octagraphene.}
\end{figure}

Now let us discuss the mechanical properties of octagraphene. As it
has a C$_{4v}$ symmetry, only three elastic constants $c_{ij}$ are
independent. Our DFT calculations estimate that the values of
$c_{11}$, $c_{12}$ and $c_{44}$ of octagraphene are 296, 46, and 49
$N/m$, respectively. The Young's modulus $E$ of octagraphene is
obtained to be 306 and 168 $N/m$ along the sawtooth and armchair
directions (see below), respectively, in comparison to those of
graphene 350 $N/m$, graphyne 245 $N/m$ and graphdiyne 174 $N/m$. The
stress-strain curves for graphene, octagraphene, graphyne and
graphdiyne under a uniaxial tension along two non-chiral directions
are illustrated in Fig. 3(a). The ultimate tensile strength or the
breaking strength $\sigma$ of octagraphene is 34.4 and 27.3 $N/m$
along the sawtooth and armchair directions, respectively, which are
quite comparable with those of graphene 38.4 $N/m$ (zigzag
direction) and 34.8 $N/m$ (armchair direction) but larger than those
of graphyne (23.4 and 19.5 $N/m$) and graphdiyne (16.8, 13.4 $N/m$).
The experimental values \cite{science_tension} of $E$ and $\sigma$
for graphene are 340 $\pm$ 50 $N/m$ and 42 $\pm$ 4 $N/m$,
respectively, showing that our DFT calculations are quite reliable.
The Poisson's ratio of octagraphene is 0.13 (sawtooth direction) and
0.47 (armchair direction), which are comparable with 0.21 (zigzag
direction) and 0.17 (armchair direction) of graphene. The Poisson's
ratio of graphyne and graphdiyne can be found in Table I. From above
results we see that octagraphene has the mechanical properties very
similar to graphene, and might be the strongest carbon atomic sheet
after graphene till now.

\begin{table*}[tbp]\footnotesize
\caption{The symmetry, lattice constant ($a_0$),  bond lengths
$l_{CC}$, plane density $\rho$, cohesive energy ($E_c$), energy gap
$E_g$ between the bottom of conduction band and the top of valence
band, Young's modulus $E$, breaking strength $\sigma$ and Poission's
ratio $\nu$ for graphene, octagraphene, graphyne and graphdiyne,
respectively.}
\begin{tabular}{ccccccccccc}
\hline\hline
Structure&method &symmetry &$a_0$ (\AA) & $l_{CC}$ (\AA) &$\rho$ ($mg/m^2$) & $E_c$ ($eV/atom$) &$E_g$ ($eV$)& $E$ ($N/m$) & $\sigma$ ($N/m$)&$\nu$\\
\hline
graphene &GGA &$C_{6v}$&2.46  &1.42&0.77&7.92&0&350&38.4,34.8&0.21,0.17\\
graphene &Exp.\cite{geim2004,possion_ratio,PR1955,andri,science_tension} &$C_{6v}$&2.458  &1.419&0.77&7.37&0&340$\pm$50&42$\pm$4&0.165\\
octagraphene&GGA & $C_{4v}$ & 3.45  &1.48,1.35&0.68&7.39&-4.03&306,168 &34.4,27.3&0.13,0.47\\
graphyne &GGA &$C_{6v}$& 6.88 &1.22,1.42&0.59&7.26&0.48&245 &23.4,19.5&0.38,0.41\\
graphdiyne &GGA& $C_{6v}$&9.45 &1.23,1.43&0.47&7.14&0.51 &174&16.8,13.4&0.41,0.44 \\
 \hline\hline
\end{tabular}
\end{table*}

\begin{figure}[tbp]
\includegraphics[width=10.5cm,clip]{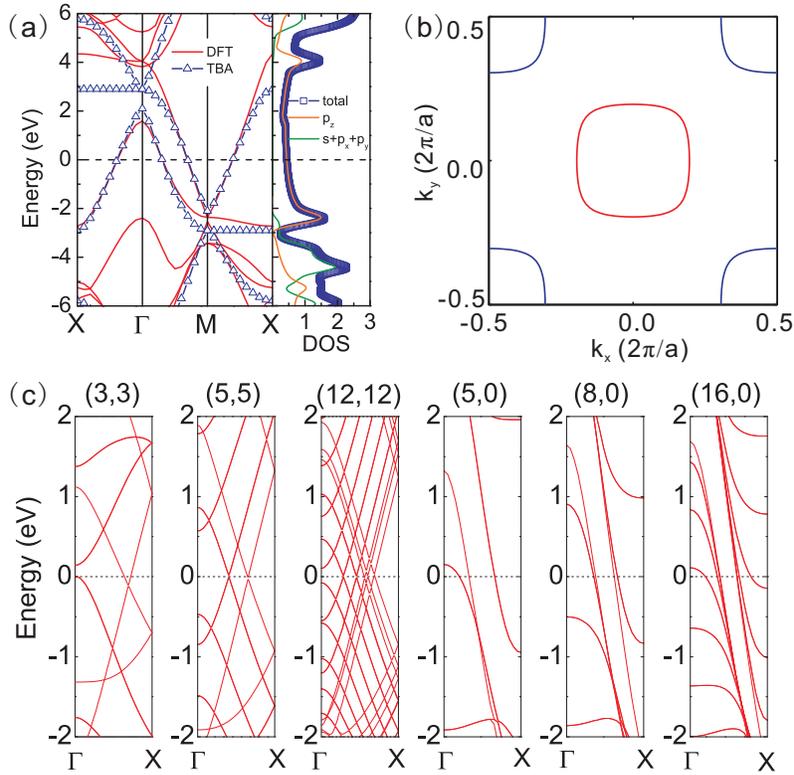}
\caption{(Color online) (a) Electronic band structures and density
of states (DOS) of octagraphene calculated by the density functional
theory (DFT) and the tight-binding approximation (TBA) methods. (b)
Fermi surface of octagraphene, where the hole packet (red) is at
$\Gamma$ point, and the electron pocket (blue) is at $M$ point. (c)
Energy bands of octagraphene nanotubes (3,3), (5,5), (12,12), (5,0),
(8,0) and (16,0).}
\end{figure}

\subsection{Electronic structures}

The valence electron density, electronic structures and the DOS of
octagraphene are obtained by means of the DFT and TBA calculations,
respectively. From Fig. 3(b) one may see that more electrons are
distributed on the inter-square bonds than on the intra-square bonds
in octagraphene, which is well consistent with its structural
feature, showing an unequal $sp^2$ hybridizing character. The
electronic structures in Fig. 4(a) show that octagraphene is a
semimetal, as its top of valence band and the bottom of conduction
band are located at $\Gamma$ and $M$ points in the Brillouin zone,
respectively, giving rise to an indirect \textit{negative} band gap
(-4.03 eV), and leading to the Fermi level passes through both the
conduction and valence bands. Thus, in contrast to graphene, the
charge carriers in octagraphene have both holes and electrons around
the Fermi level. Note that graphene is also a semimetal or a
zero-gap semiconductor because its valence and conduction bands
touch at $K$ points, while graphyne and graphdiyne are
semiconductors with narrow band gap of 0.48 and 0.51 $eV$,
respectively.

The projected DOS of octagraphene shows that the energy bands near
the Fermi surface are predominantly contributed by the $2p_z$
orbital of carbon atom, forming the same $\pi$-bond as that in
graphene. The Fermi surface of octagraphene is plotted in Fig. 4(b),
which consists of a hole pocket at $\Gamma$ point and an electron
pocket at $M$ point. Since each unit cell contributes four electrons
with two bonding states and two antibonding states, each point in
the Brillouin zone should be filled with $2\times 2$ electrons, and
the number of unoccupied states in valence band is just equal to
that of occupied states in conduction band, resulting in that the
hole and electron pockets share the same area.

\subsection{Low-energy physics}

To describe the low-energy physics near the Fermi surface of
octagraphene, we propose a tight-binding model which contains only
$\pi$ electrons with nearest neighbor hoppings. The Hamiltonian can
be written as
\begin{equation}
H = t \sum_{\langle i,j \rangle,\sigma}
(\hat{c}^{\dagger}_{i,\sigma} \hat{c}_{j,\sigma} + h.c.) + t'
\sum_{\ll i,j \gg,\sigma} (\hat{c}^{\dagger}_{i,\sigma}
\hat{c}_{j,\sigma} + h.c.),
\end{equation}
where $\hat{c}_{i,\sigma}$ and $\hat{c}^{\dagger}_{i,\sigma}$ are
the electron annihilation and creation operators at site $i$ with
spin $\sigma$, $\langle i,j \rangle$ runs over the nearest neighbors
within the same square and $t$ is the corresponding hopping
amplitude of electrons, $\ll i,j \gg$ runs over the nearest
neighbors connecting different squares and $t'$ is the corresponding
hopping amplitude. In the matrix form, the Hamiltonian reads
\begin{equation}
H_{\textbf{k}} = \left(
\begin{array}{cccc}
0 & t & t^{\prime }e^{-ik_{y}} & t \\
t & 0 & t & t^{\prime }e^{ik_{x}} \\
t^{\prime }e^{ik_{y}} & t & 0 & t \\
t & t^{\prime }e^{-ik_{x}} & t & 0%
\end{array}
\right), \label{matrix}
\end{equation}
where $k_{x}$ and $k_{y}$ are $x$ and $y$ components of momentum of
$\pi$ electrons, and the lattice spacing $a_0=1$ is taken. The
dispersion $E(\textbf{k})$ of octagraphene within the TBA is
determined by
\begin{eqnarray}
E(\textbf{k})^{4}&-& 2(t'^{2}+2t^{2})E(\textbf{k})^{2}+4t't^{2}(\cos
k_x + \cos k_y)E(\textbf{k}) \nonumber \\
&+& t'^{2}[t'^{2}-4t^{2}(\cos k_x + \cos k_y)]=0. \label{dispersion}
\end{eqnarray}
The TBA results are in good agreement with those obtained by the DFT
calculations [Fig. 4(a)], implying that the low-energy properties of
$\pi$ electrons in octagraphene can be well described by the TBA
Hamiltonian, where the fittings give $t=-2.5$ $eV$ and $t'=-2.9$
$eV$. For graphene, $t=-2.7$ $eV$ \cite{grapheneTBA}.

\subsection{Fabricating suggestions}

To obtain the structure of octagraphene experimentally, we suggest
that one may make the line defects periodically in graphene, and
then connect properly the broken bonds of carbon atoms, as
demonstrated in Fig. 5. In addition, it may be possible to
synthesize it through the acetylene scaffolding ways
\cite{chemsocrev,chemrev}.

\section{octagraphene nanotubes and fullerenes}

By rolling octagraphene, the novel single-walled carbon nanotubes
can also be obtained. For this purpose, let us define the chiral
vector $\vec{C}_h$ of octagraphene carbon nanotube (OCNT) by
$\vec{C}_h=n\vec{a}_1 + m\vec{a}_2 \equiv (n,m)$, where $n$ and $m$
are integers with $0\leq |m| \leq n$ because of the $C_{4v}$
symmetry. We call $(n,0)$ the sawtooth OCNT ($m=0$,
$\theta=0^{\circ}$), and $(n,n)$ the armchair OCNT ($m=n$,
$\theta=45^{\circ}$), where the chiral angle is defined as $\cos
\theta =n/\sqrt{n^2 + m^2}$. The tube diameter is given by $d_t =
|\vec{C}_h|/\pi = a_0 \sqrt{n^2+m^2}/\pi$. The geometric structures
of sawtooth OCNT (10,0) and armchair OCNT (8,8) are presented in
Fig. 1(b) as examples, whose structural stabilities were verified by
the DFT calculations. The energy bands of OCNT (3,3), (5,5),
(12,12), (5,0), (8,0) and (16,0) are given in Fig. 4(c), which show
that regardless of the curvature effect, the single-walled OCNTs,
either sawtooth or armchair, are all metallic.

In addition, by wrapping octagraphene and using pentagons, hexagons
or heptagons as caps, many structurally stable unconventional carbon
fullerenes can be built. The optimized examples like C$_{36}$,
C$_{48}$, C$_{80}$ and C$_{96}$ are shown in Fig. 1(c). These
results show that octagraphene is indeed a versatile 2D crystalline
allotrope of carbon after graphene.

\begin{figure}[tbp]
\includegraphics[width=1.0\linewidth,clip]{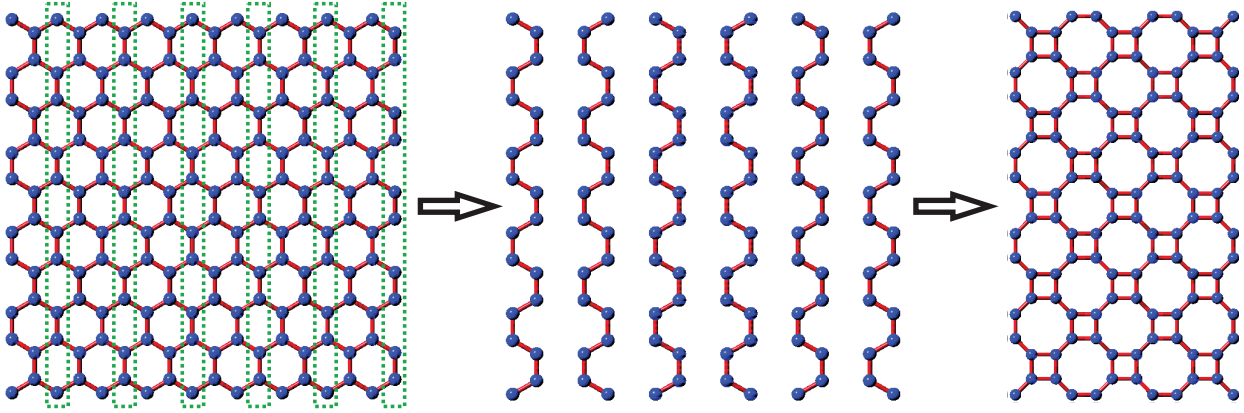}
\caption{(Color online) A schematic demonstration how to utilize the
line defects in graphene to obtain octagraphene.}
\end{figure}

\section{Band engineering and hydrogen storage}

\begin{figure}[tbp]
\includegraphics[width=8.5cm]{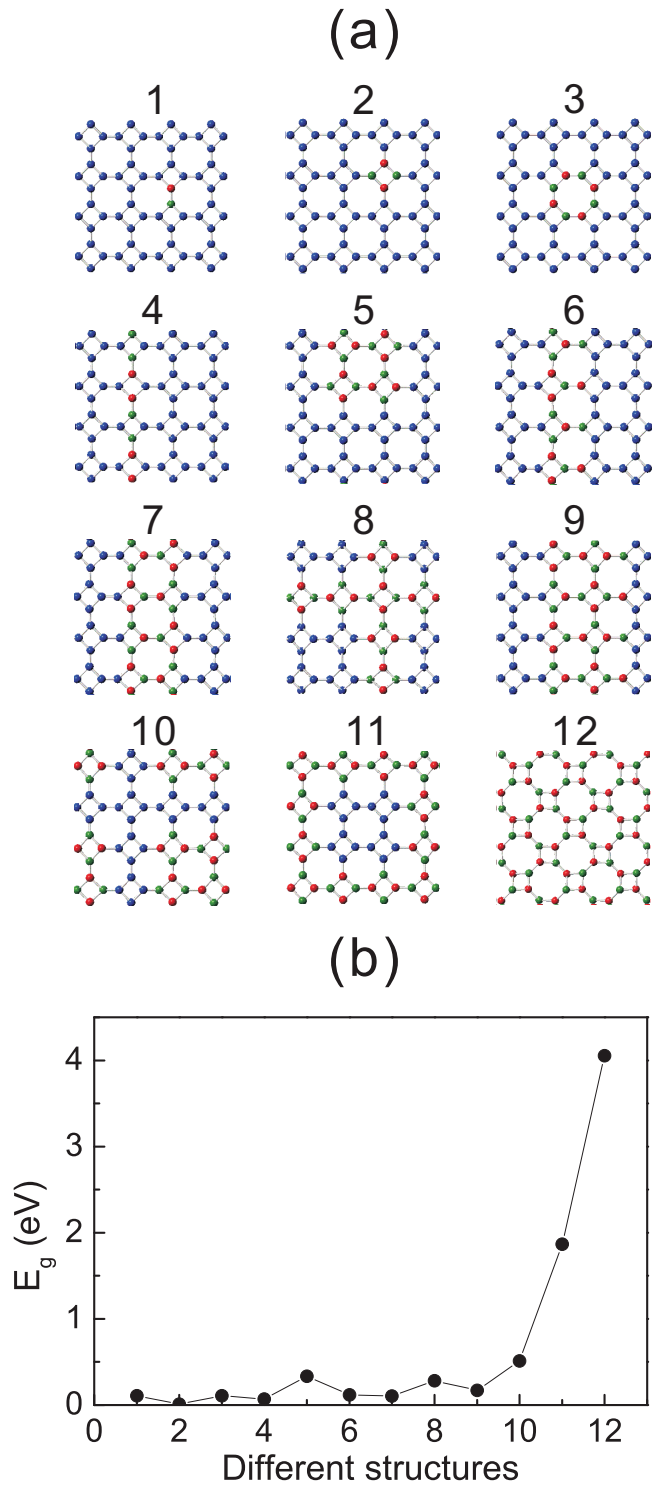}
\caption{(a) Schematic structures of boron nitrogen pairs
substitutionally doped octagraphene with different doping
configurations, where carbon, boron, and nitrogen atoms are colored
in blue, red, and green, respectively. (b) The corresponding band
gaps (E$_g$) for different structures of (a), where the doping
concentration is from small to large in sequence.} \label{gap}
\end{figure}

\subsection{Boron nitrogen pair doping}

Similar to graphene and carbon nanotubes\cite{graphene_dop,cnt_dop},
the semimetallic octagraphene may be applicable in the band
engineering via the chemical doping. In fact our calculation reveals
that a band gap opens if the octagraphene is substitutionally doped
with boron and nitrogen (B-N) pairs, as shown in Fig. \ref{gap},
where eleven structures for different doping configurations and a
pure B-N octa-square structure are considered. All these doped
structures are energetically favorable. In Fig. \ref{gap} (a), the
structures of number 1-11 are those of octagraphene of 4$\times$4
supercell with substituting carbon (blue ball) by boron (red ball)
and nitrogen (green ball) atom pairs, and the structure of number 12
is the pure B-N octagon-square structure of 3$\times$3 supercell. In
the first structure two carbon neighbors are replaced by one B-N
pair; in the second (third) structure a square (octagon) is
substituted by two (four) B-N pairs; in the fourth (sixth, seventh
or ninth) one a nanoribbon in octagraphene is substituted by four
(eight, twelve or sixteen) B-N pairs; in the fifth one a 2$\times$2
supercell is replaced by eight B-N pairs; in the eighth one a cross
is replaced by fourteen B-N pairs; the tenth one is the dual
structure of the eighth; the eleventh is the dual structure of the
fifth. The band gap varies significantly for different doping
configurations, as shown in Fig. \ref{gap} (b). It is seen that the
boron nitrogen pair doped octagraphene can open a gap from 0.009 eV
(structure 2) to 1.866 eV (structure 11), which can gain some
insights for nanoelectronics. It is noting that the band gaps of
most of structures considered here are around 0.2 eV, even when the
doping concentration is as large as 50.82\% (structure 9), which is
quite different from the doped graphene \cite{Shinde}. Considering
that the GGA usually underestimates the band gap, the present doping
study gives useful information for the band engineering in
octagraphene.

\begin{figure}[tbp]
\includegraphics[width=5.5cm]{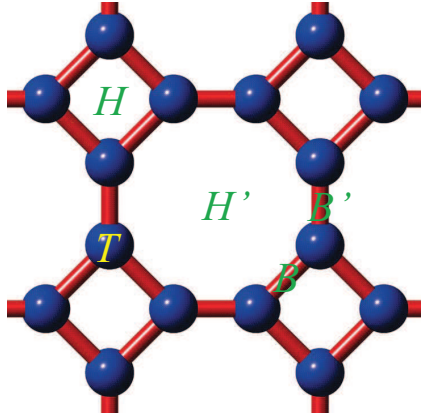}
\caption{(Color online) There are five possible adsorption sites of
metal atoms on octagraphene, including one top site on C atoms
($T$), two bridge sites on top of intra-square ($B$) and
inter-square ($B'$) C-C bond, respectively, and two hollow sites on
top of square ($H$) and
  octagon ($H'$), respectively.}
\label{site}
\end{figure}

\begin{figure}[tbp]
\includegraphics[width=9.5cm]{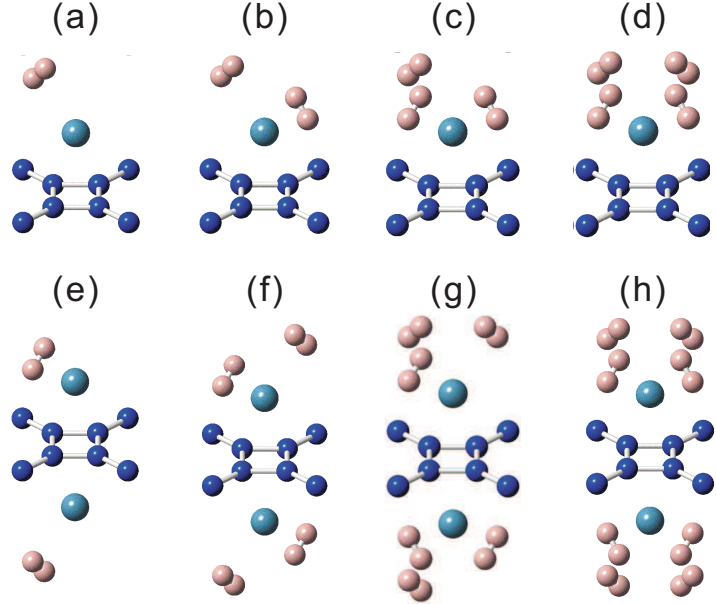}
\caption{(Color online) The configuration of four hydrogen molecules
  adsorbed on each Ti atom at H$_{1}$ site. (a)-(d) reveal the single-side
  hydrogen storage configurations with one, two, three and four hydrogen
  molecules adsorbed on each Ti atom, respectively; (e)-(h) show the double-side
  hydrogen storage configurations with one, two, three and four H$_2$ adsorbed
  on each Ti atom, respectively; where C, Ti and H atoms are colored in dark blue,
  bluish green and brick red, respectively.}
\label{Hydrogen}
\end{figure}

\subsection{Hydrogen storage}

Next, let us consider the possibility of hydrogen storage in
octagraphene. In general, there are two ways to realize this
purpose. One is that hydrogen atoms can be absorbed on carbons via
the chemical bond to form a new hydrocarbon sheet, similar to
graphane \cite{graphanePRB}; the other is that the hydrogen
molecules can be adsorbed physically on some metal atoms (e.g. Ti,
Ca, Al, etc.) that are deposited on the sheet of octagraphene. In
the former case, the hydrogen storage capacity of octagraphene can
reach 7.7 wt\% as graphane does \cite{graphanePRB}. In the latter
case, the hydrogen capacity depends on the adsorption configuration
of metal atoms. There are five possible adsorption positions for
metal atoms (see Fig. \ref{site}), including one top site on C atoms
($T$), two bridge sites on top of intra-square ($B$) and
inter-square ($B'$) C-C bond, respectively, and two hollow sites on
top of square ($H$) and octagon ($H'$), respectively. To analyze the
stability of these five adsorption configurations, we calculate the
cohesive energy $E_{b-M}$ with $M$ the adsorbed metal atom by
\begin{eqnarray} \label{eq:1}
E_{b-M}=[E_{octa}+nE_{M}-E_{M-octa}]/n,
\end{eqnarray}
where $E_{octa}$, $E_{M}$ and $E_{M-octa}$ are the total energies of
pure octagraphene, $M$ atom and octagraphene with adsorbed metal
$M$, respectively. For Ti doping ($M$=Ti), $H$ site is the most
stable position with cohesive energy of 2.71 eV per Ti atom.

Figs. \ref{Hydrogen}(a)-(d) present single-side hydrogen storage
configurations with one, two, three and four hydrogen molecules
adsorbed on each Ti atom. The average adsorption energy per H$_2$ is
0.207 eV, 0.361 eV, 0.336 eV and 0.325 eV, respectively. Figs.
\ref{Hydrogen} (e)-(h) give double-side hydrogen storage
configurations with one, two, three and four H$_2$ adsorbed on each
Ti atom. The average adsorption energy per H$_2$ is 0.235 eV, 0.355
eV, 0.336 eV and 0.325 eV, respectively. In this case, the hydrogen
storage capacity could reach 7.76 wt\% for Ti [as shown in Fig.
\ref{Hydrogen}(h)], which is very close to that of Ca-adsorbed
graphene, say 8.4 wt\% \cite{grapheneH}. We also perform the same
calculation for Al and Ca atoms ($M$=Al or Ca), respectively, and
found that the adsorption energy of hydrogen molecule is too small.
The adsorption energy of Ti-adsorbed octagraphene is quite
promising, say, about 0.3 eV/H$_2$. It appears that the hydrogen
storage capacity of octagraphene is close to that of fullerenes
\cite{C60H}, carbon nanotubes \cite{cntH} and graphene
\cite{grapheneH}.

\section{Conclusion}

In conclusion, we propose a periodic carbon sheet by the
first-principles calculations. This novel carbon monolayer is a
$sp^2$-hybridized structure and coined as octagraphene. Unlike
graphene, octagraphene is composed of carbon squares and octagons
possessing $C_{4v}$ symmetry, and is more favorable in energy than
graphyne and graphdiyne. Its optimized structure has two bond
lengths with 1.48 {\AA} within each square and 1.35 {\AA} between
adjacent squares. The band structure shows that it is a semimetal
with Fermi surface consisting of one hole pocket and one electron
pocket. In order to describe the low-energy physics of octagraphene,
we also suggest a tight binding model of $\pi$ electrons which is
uncovered to agree well with DFT calculations, and is quite helpful
for further exploring the properties of octagraphene in a magnetic
field or emergent quantum phenomena such as quantum Hall effect,
Kondo effect, and so on in octagraphene. The intriguing mechanical
properties of octagraphene are also obtained, which are observed
close to those of graphene. Similar to graphene, we find that
octagraphene can also be rolled seamlessly to form stable carbon
nanotubes and unconventional fullerenes. By substitutionally doping
boron nitrogen pairs, we disclose that the semimetallic octagraphene
can be turned into a semiconductor with a band gap, the property
will be useful for possible applications in nanoelectronics. This
new carbon allotrope has also the possibility for hydrogen storage.
Our calculation shows that for the 2D octagraphene sheet the
hydrogen storage capacity could reach 7.76\% via Ti-adsorption that
is close to that of graphene. Finally, we would like to mention that
one might obtain this new kind of carbon atomic sheet via making
line defects in graphene or through the acetylene scaffolding ways.

\acknowledgments

We are grateful to Dr. Eric Germaneau for useful discussions. All
calculations were completed in the Shanghai Supercomputer Center,
China. This work is supported by the NSFC (Grants No. 90922033, No.
10934008, No. 10974253, No. 10904081 and No. 11004239), the MOST of
China (Grant No. 2012CB932900) and the CAS.

\end{document}